\documentstyle[prb,aps,multicol,12pt,epsf]{revtex}

\author{S. N. Dorogovtsev\cite{E}}
\title{Avalanche mixing of granular solids in a rotating 2D drum: \\
diffusion and fractionality
}
\draft
\address{A.F.Ioffe Physico-Technical Institute, 194021 St.Petersburg, Russia
}

\begin{document}

\maketitle
\begin{abstract}
The dynamics of the avalanche mixing
in a slowly rotated 2D upright drum is studied in the situation where
the difference $\delta$ between the angle of marginal stability
and the angle of repose of the granular material is finite.
An analytical solution of the problem is found for a half filled drum,
that is the most interesting case.
The mixing is described by a simple linear difference equation.
We show that the mixing looks like
linear diffusion of fractions under consideration with the
diffusion coefficient vanishing
when $\delta$ is an integer part of $\pi$.
The characteristic mixing time tends to infinity in these points.  A full
dependence of the mixing time on $\delta$ is calculated and predictions
for an experiment are made.
\end{abstract}



The avalanche mixing problem \cite{metcal,dor1,peratt1,dor2,peratt2},
may be, is the only one among all numerous problems concerning
the evolution of granular solids in rotating drums
\cite{po,ris1,raj,baum,zik,clem,can2} that may be formulated
with absolute mathematical clearness \cite{dor1,peratt1}.

\begin{figure}[\!h]
\epsfxsize=3.5in
\epsffile[-128 233 257 569]{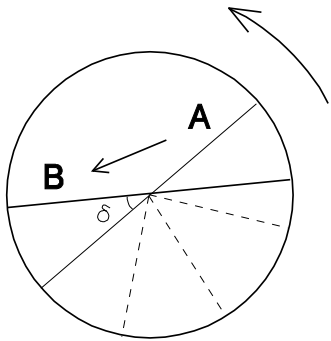 }
\caption{
\narrowtext
The avalanche mixing scheme in the case of a half filled drum.
When the free surface slope
attains the marginal stability angle value, the granules
of different fractions flow quickly from sector $A$ to sector $B$,
undergoing mixing. Then the free surface slope turns to be equal to
the repose angle of the material under consideration.
The angle $\delta$ is the difference between the angle of marginal
stability and the angle of repose. The mixing fractions are not shown.
\label{fig1}}
\end{figure}

Let us define first the so called avalanche mixing \cite{metcal}.
A partially filled 2D upright drum rotates slowly (see Fig.~\ref{fig1}).
One can assume its radius be equal to $1$, the angle of rotation
of the drum plays the role of time. The free surface slope
varies periodically between the angle of repose
and the angle of marginal stability of the material.
Small granules of different fractions (they differ only by color)
move as a whole with a drum with one exception: periodically,
when the free surface slope achieves the marginal stability value,
the granules from the sector of the angle $\delta$ near the upper half of
a free surface fall down (see Fig.~\ref{fig1} -- granules fall down
from sector $A$ to sector $B$). Here $\delta$ is the difference
between the angle of marginal stability and the angle of repose.
Therefore, the only reason for mixing in the problem is granular
flow (avalanches) between the wedges.
One should describe, how the mixing develops in time.

We assume that the fractions becomes be mixed homogeneously after
the fall -- the mixture in sector $B$ in Fig.~\ref{fig1}. This case,
in which the avalanche mixing proceeds most quickly,
appears to be near the experimentally realized situation \cite{metcal}.
When $\delta \to 0$, the mixing can be described in the frames of a simple
geometrical approach \cite{metcal,dor1,peratt1,dor2,peratt2}, and
an analytical theory \cite{dor1,dor2} can be used.
There is no such clearness, when $\delta$ is finite \cite{peratt1,peratt2}.
Below we consider analytically the finite $\delta$ case in the most
interesting situation of a half filled drum. (When a drum is half filled
and $\delta=0$, the mixing is absent --
see papers \cite{metcal,dor1,peratt1,dor2,peratt2},
so in this situation, the effect of nonzero $\delta$ is most striking.)

Let us assume that there are two fractions -- black and white. Then
the mixing dynamics is described by a set $\{c_m\}$. Here $c_m$ is
the concentration of the black fraction in sector $B$ (Fig.~\ref{fig1})
after the $m$-th avalanche falls, that will be at the moment
$t=m\,\pi/\delta$. In the case of a half filled drum, the equation
for $c_m$ appears to be very simple:

\begin{equation}
\label{L1}
c_m=
\left\{1- \left(\frac\pi\delta-\left[\frac\pi\delta\right]\right) \right\}
c_{m-[\pi/\delta] }+
\left\{\frac\pi\delta
- \left[\frac\pi\delta\right] \right\} c_{m-[\pi/\delta]-1}
\end{equation}
(it can be understood easily from Fig.~\ref{fig1}).
Here $[ \cdot ]$ denotes an integer part of a number.
The values $c_m, \, m=-[\pi/\delta],-[\pi/\delta]+1,\ldots,-1,0$
are used as an initial condition. We shall use the notation
$n \equiv [\pi/\delta]$ for the integer part and
$a \equiv \pi/\delta-[\pi/\delta]$ for the fractional part.

It is not a simple task to write directly
an explicit form of the solution of Eq.~(\ref{L1}), if $n>2$.
Let us find the solution in the case where $\delta \ll \pi$, i.e. $n \gg 1$.
The idea of the trick to be used is obvious from the following
form of Eq.~(\ref{L1}):

\begin{equation}
\label{L2}
c_{m+n}-c_{m}=-a(c_m-c_{m-1}) \ .
\end{equation}
One is tempted to introduce different variables for the different scales:
a la time variable $\tilde{t}$ for a large scale in the left-hand side of
Eq.~(\ref{L2}) and an angle variable $\varphi$ for a short scale in the
right-hand side.

We shall pass on carefully to continual variables
(the necessary notions will be made later).
First, let $m$ be continual now.
Then $\varphi=m\delta$. We shall define the variable $\tilde{t}$ in the
following way. Let us calculate the following difference:
$c_{m+n}-c_{m-q(a)}$, where $q(a)$ is unknown function for the moment.

\begin{eqnarray}
\label{L3}
c_{m+n}-c_{m-q(a)}=-a(c_{m}-c_{m-1})+c_{m}-c_{m-q(a)} \to
\phantom{eeeeeee}\nonumber \\
\delta(q(a)-a)\frac{\partial c(\varphi)}{\partial\varphi} +
\frac{\delta^2}{2} \left\{ (1-a)q^2 (a) +a(1-q(a))^2 \right\}
\frac{\partial^2 c(\varphi)}{\partial\varphi^2}  \ .
\end{eqnarray}
Now we can choose $q(a)$ demanding the coefficient of the first derivative
be zero. Therefore, $q(a)=a$ and one may write

\begin{equation}
\label{L4}
c_{m+n}-c_{m-a} \to (n\delta+a\delta)\frac{\partial c}{\partial\tilde{t}}=
\pi \frac{\partial c}{\partial\tilde{t}} \ .
\end{equation}
Hear the time variable $\tilde{t}$ is introduced in the following way.
We observe periodically the system under consideration
with the periodicity interval
$\pi$ (like using the stroboscopic effect) and describe the result of
the observation by the continual variable $\tilde{t}$.

Finally, we get the partial differential equation
\begin{equation}
\label{L5}
\frac{\partial c(\varphi,\tilde{t})}{\partial\tilde{t}}=
\frac{\delta^2 a(1-a)}{2\pi}\,
\frac{\partial^2 c(\varphi,\tilde{t})}{\partial\varphi^2} \ .
\end{equation}
One can easily verify that the boundary conditions for Eq.~(\ref{L5})
are $c(\varphi=0)=c(\varphi=\pi)$ and
$\partial c(\varphi=0)/\partial\varphi=
\partial c(\varphi=\pi)/\partial\varphi$. The initial distribution of the
fractions defines the initial condition for Eq.~(\ref{L5}).
Thus, the avalanche mixing is described by linear diffusion equation
(\ref{L5}) with the diffusion
coefficient that vanishes at the angles $\delta=\pi/k$, where $k$ is
integer.
Note, that previously, the resonance-like behavior of some other
quantity due to the same reason was found numerically
\cite{peratt1,peratt2}.

The solution $c(\varphi,\tilde{t})$ of Eq.~(\ref{L5}) solves the problem,
since, of course, one can find trivially all $c_m$ from
$c(\varphi,\tilde{t})$. We shall not present this expression here but
shall write immediately the answer for the mixing starting from the
following initial configuration. Let it be a thin layer of the black
fraction over the white fraction in the drum at the initial moment.
Then, in our situation, one may use the initial condition
$c(\varphi,\tilde{t}=0)=4\gamma\delta(\varphi)$. Here $\delta(\varphi)$
is the $\delta$-function and $4\gamma$ defines the amount of the black
fraction. Then the solution is of the form

\begin{eqnarray}
\label{L6}
c(\varphi,\tilde{t})=4\gamma \left\{\frac{1}{\pi}+
\frac2\pi \sum_{k=1}^{\infty}
\exp\left\{ -k^2 \frac2{\pi}\,\delta^2 a(1-a) \tilde{t} \right\}
\cos\left( 2k \varphi \right)\right\}=
\nonumber \\
\frac{4\gamma}{2\pi} \theta_3 \left( \frac{\varphi}{\pi},
\frac{2}{\pi^3}\,\delta^2 a(1-a) \tilde{t} \right) \ ,
\phantom{eeeeeeeeeeeeeeeee} \
\end{eqnarray}
where $\theta_3$ is the theta-function.
If $\tilde{t} \ll \pi/(2\delta^2 a(1-a))$,

\begin{equation}
\label{L7}
c(\varphi,\tilde{t})=\frac{4\gamma}{\delta\sqrt{2a(1-a)\tilde{t}}}
\exp\left\{-\frac{\varphi^2}{2\delta^2a(1-a)\tilde{t}/\pi} \right\}  \ .
\end{equation}
For longer time one obtains exponential relaxation with
the relaxation time $\tau$ (i.e. the mixing time):

\begin{equation}
\label{L8}
\tau^{-1}= \frac{2}{\pi} \delta^2 \left(\frac\pi\delta-
\left[\frac\pi\delta\right]\right)
\left\{1- \left(\frac\pi\delta-\left[\frac\pi\delta\right]\right) \right\}
\end{equation}
to a homogeneous mixture $c_\infty=4\gamma/\pi$.
(Of course, expression (\ref{L8}) is valid for any initial conditions.)

Note that continualization of difference equations is very dangerous
trick (e.g. see the discussion of the topic in the handbook of E.~Pinney
\cite{pinney}). Nevertheless, in our specific case, it works!
We shall not try to justify it rigorously,
(one of the reasons of its availability
is fast decreasing of all derivatives of
$c(\varphi,\tilde{t})$ and the other one is the separation of the scales --
see Eq.~(\ref{L2})) but shall compare the
main derived result, i.e. the mixing time, with that obtained in
a more direct way.

According to usual prescriptions (e.g. see handbook \cite{pinney}),
one may search for the solution of Eq.~(\ref{L1}) or Eq.~(\ref{L2})
in the form of
a linear combination of the terms $\lambda_j^m$, where $\lambda_j$
are the roots of the characteristic polynomial

\begin{equation}
\label{L9}
\lambda^{n+1}-(1-a)\lambda-a=0 \ .
\end{equation}
\vspace{.3in}
\begin{figure}[\!h]
\epsfxsize=3.5in
\epsffile{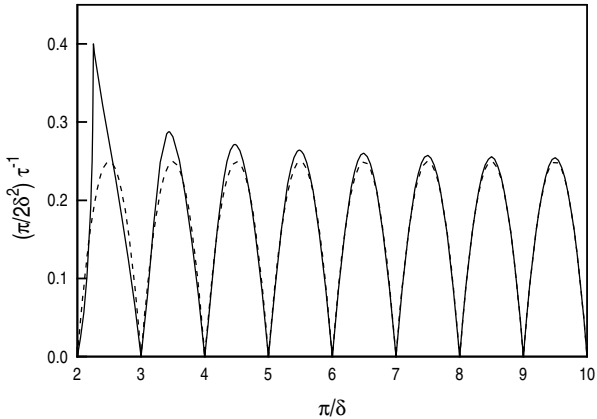 }
\caption{
\narrowtext
Scaled inverse avalanche mixing time $(\pi/2\delta^2)\tau^{-1}$ vs.
$\pi/\delta$.
The solid line is obtained from the "exact" solution of the characteristic
polynomial (\protect\ref{L9}).
The dashed line represents expression (\protect\ref{L8})
which is, formally speaking, obtained in the limit
of small $\delta \ll \pi$.
\label{fig2}}
\end{figure}
\noindent
If $\mu_j$ is the multiplicity of the $\lambda_j$ root, then the
corresponding contributions to the general solution of Eq.~(\ref{L2}) are
$m^{s_j}|\lambda_j|^m \cos(\omega_jm),
\,m^{s_j}|\lambda_j|^m \sin(\omega_jm)$, where $s_j=0,1,\ldots,\mu_j-1$ and
$\omega_j=arg(\lambda_j)$. Thus, to obtain the long time relaxation we should
find the maximal $|\lambda_j| \equiv \lambda_{max}$ which is lower then the
root $\lambda=1$ of Eq.~(\ref{L9}). The mixing time is expressed in the
terms of $\lambda_{max}$:
$\tau^{-1}=(1/\delta)\log(1/\lambda_{max})$,
so it may be calculated directly for any value of $a$ and a finite $n$.
The result is shown by the solid line in Fig.~\ref{fig2}.
The dashed line represents the answer (\ref{L8}) obtained for small
angles $\delta \ll \pi$.
The closeness of these curves (even for large angles $\delta$) looks
impressing. It verifies the quality of the method used above.

In an experiment, the angle $\delta$ usually fluctuates with time.
One can see easily that if the deviations from the
mean value of $\delta$ are greater then a very small angle
$\delta^2/\pi$ $(\delta/\pi \ll 1)$, then they should smear the dependence
$\tau(\delta)$ and eliminate the singularities.
It is not an easy task to take into account exactly these fluctuations.
The most naive estimation for the averaged value looks like
$\tau^{-1} \sim \delta^2/3\pi$. In the experiment \cite{metcal},
$\delta=8^\circ \pm 2^\circ$, so from our estimation, $\tau \sim 80$ turns
of the drum. Unfortunately, we did not find any experimental data on this
quantity for a half filled drum, so the comparison is impossible at the
moment.

In conclusion, we have described analytically the dynamics of the avalanche
mixing in the case of a half filled drum where the difference between
the angle of the marginal stability and the angle of repose plays the
principal role. The mixing looks like diffusion with the diffusion
coefficient vanishing when $\delta$ is an integer part of $\pi$.

I wish to thank V. V. Bryksin, Yu. A. Firsov,
A.~V.~Goltsev,  S.~A.~Ktitorov, E.~K.~Kudinov, B.~N.~Shalaev
for helpful discussions and B.~A.~Peratt and J.~A.~Yorke for sending
their paper \cite{peratt2} before the publication.
This work was supported in part by the RFBR grant.

\end{document}